\documentclass[11pt]{iopart}
\usepackage{graphicx}
\usepackage{amsfonts}		
\expandafter\let\csname equation*\endcsname\relax        
\expandafter\let\csname endequation*\endcsname\relax
\usepackage{amsmath}
\usepackage{color}

\begin{document}
\title{Wrapping probabilities for Potts spin clusters on a torus}
\author{T. Blanchard}
\address{Sorbonne Universit\'es, UPMC Univ Paris 06, UMR 7589, LPTHE, F-75005, Paris, France}
\address{CNRS, UMR 7589, LPTHE, F-75005, Paris, France}
\ead{blanchard@lpthe.jussieu.fr}

\date{\today}

\begin{abstract}
\noindent 
In this work we obtain wrapping probabilities for Ising spin clusters on a torus.
We use the analogy with the tricritical point of a Potts model with dilution. The formula obtained are tested against
numerical simulation. We also provide closed form for the wrapping probabilities of cross clusters on untwisted tori. 
\end{abstract}


\maketitle

\section*{Introduction}
In the study of the critical properties of the Potts model, it was soon remarked that rewriting the model in terms of
clusters was very enlightening~\cite{fortuin_random-cluster_1972,coniglio_clusters_1980}. Indeed, the Fortuin-Kasteleyn
(FK) clusters or droplets, obtained by casting the Potts partition function into a random cluster partition function,
contain all the critical properties of the model and in particular are ruled by the same critical exponents. However,
these are not the clusters one would observe in an experimental realisation of the Potts model, in this case one can
only see the clusters formed by connected spins sharing the same value (spin clusters). In two dimensions both type of
clusters undergo a continuous percolation transition at the critical temperature but a measure of the fractal dimensions
of FK and spin clusters quickly convinces that the two types of clusters possess different critical
exponents~\cite{janke_fractal_2005}. From a theoretical point of view there is then the question of the origin of this
set of exponents ruling the properties of spin clusters. A clear answer can be given for the $Q=2$ states Potts model,
\textit{i. e.} the Ising model. In~\cite{coniglio_clusters_1980} it was shown that the Ising Hamiltonian can be
rewritten as a diluted Potts model with $Q'=1$ state. This enabled the computation of the Ising spin clusters
exponents~\cite{stella_scaling_1989}. Unfortunately no such mapping exists for other Potts models. It was nonetheless
conjectured in~\cite{vanderzande_fractal_1992} that the exponents for spin clusters can be obtained by a similar mapping
to a diluted Potts model with a different number of states at its tricritical point. This conjecture gives exponents
compatible with numerical estimates~\cite{wu_potts_1982,janke_geometrical_2004,deng_geometric_2004} and there is little
doubt that those are exact even though a rigorous proof is still lacking.  The situation was thought to be more or less
under control until recent works showed that a naive analytic continuation from the critical branch sometimes fails to give the
correct results for some spin clusters properties. Even though no problem is encountered at the level of dimensions,
a naive continuation is not able to explain the connectivity constants appearing in bulk three-point
connectivity~\cite{delfino_spin_2013} and the corner contribution to cluster
number~\cite{kovacs_corner_2013}. It is thus very interesting to further study the properties of spin clusters to
understand what can and cannot be obtained from an analytic continuation.

In percolation theory, it is a basic fact that below the percolation threshold no infinite cluster exists and that above
there exists one with probability one~\cite{stauffer_introduction_1994}. At the percolation threshold, an important
question for finite-size systems is whether a cluster spanning the whole system is present or in other words: what is the
crossing probability of a percolation cluster for a given system?
The study of crossing probabilities has a long history and a landmark in this topic is due to Cardy and his eponymous
formula~\cite{cardy_critical_1992}. This formula gives the probability of left-right crossing of an incipient spanning
cluster at the critical percolation threshold on a rectangle. This probability depends only upon the aspect ratio $r$ of
the rectangle and can be expressed in terms of hypergeometric functions.  At the same time Langlands \textit{et
al.}~\cite{langlands_universality_1992} tested numerically the universality of crossing probabilities in different
setups.  Since then a number of percolation probabilities have been studied either for
percolation~\cite{pinson_critical_1994,watts_crossing_1996,cardy_crossing_2002} or other
models~\cite{lapalme_crossing_2001,arguin_non-unitary_2002,bauer_multiple_2005,kozdron_using_2009}. One aspect that is
particularly interesting is that although Cardy and others found those formulas using conformal field theory (CFT),
\textit{i. e.} in a non rigorous way mathematically speaking, rigorous tools have been developed which can be used to tackle those
systems. Schramm introduced the stochastic Loewner evolution (SLE)~\cite{schramm_scaling_2000} which describes numerous
physically occurring curves (see~\cite{lawler_scaling_2002} with Lawler and Werner \textit{e.g.}). In a slightly
different line of research Smirnov proved
rigorously Cardy's formula and the existence of a conformal scaling limit of the percolation critical point for site percolation on the triangular
lattice~\cite{smirnov_critical_2001}. 

As a generalisation of the case of simply connected surfaces, Langlands~\textit{et al.} introduced the notion of crossing
probabilities on compact Riemann surfaces~\cite{Langlands_conformal_1994}. For percolation on a compact Riemann surface S,
one can consider the topological space $X_s$ generated by a given configuration $s$ and the homomorphism $\phi:H_1(X_s)\to
H_1(S)$ from the first homology group of $X_s$ to that of $S$. The probability $\pi_G$ that a configuration $X_s$
contains in its image by $\phi$ a subgroup $G\subset H_1(S)$ is:
\begin{equation}
\pi_G=\frac{Z_G}{Z}
\label{prob}
\end{equation}
where $Z$ is the total partition function of the model considered and $Z_G$ the partition function restricted to the
configurations generating $G$ by $\phi$.  Among other cases Langlands~\textit{et al.} studied numerically those
probabilities for percolation on a torus. Shortly after they were derived analytically for percolation by
Pinson~\cite{pinson_critical_1994} building on the work of Di~Francesco~\textit{et al.}~\cite{francesco_relations_1987}.
Arguin~\cite{arguin_homology_2002} then extended the work of Pinson to the case of FK clusters of the $Q$ states Potts
model for $Q\in [1,4]$. He was also able to derive closed forms in terms of Jacobi $\theta$ functions in the case of
percolation ($Q=1$) and Ising FK clusters ($Q=2$) and supported his formula for integer values of $Q$ with simulations.
In~\cite{morin-duchesne_critical_2009} it was checked numerically that the formula obtained by
Arguin is also valid for real values of $Q$.

In the present work we further generalise the derivation of wrapping probabilities of FK clusters by Arguin to the case
of Potts spin clusters. In the present work only the Ising model is treated. In this case a closed form in terms of
Jacobi $\theta$ function is obtained and checked numerically. Even though other cases (mainly the 3-Potts model) are not
treated explicity in this work it is expected that the arguments developped for the Ising spin clusters can be
generalized to other cases. This work was motivated by the study of the appearance of long-lived stripe states in the
Ising model after a quench from $T=T_c$ to $T=0$ with various boundary conditions~\cite{blanchard_frozen_2013}. When
periodic boundary condition are applied, the system has the topology of a torus and wrapping probabilities of spin
clusters are relevant to explain the appearance of long-lived stripe
states~\cite{barros_freezing_2009,olejarz_fate_2012}. We also present expressions for the probabilities of
cross-shaped FK and spin clusters for the Ising model on untwisted tori.  The article is divided as follows: we further
discuss FK and spin clusters, present how wrapping probabilities can be computed, show the derivation for Ising spin
clusters and check it numerically and finally we provide expressions for cross-shaped clusters.

\section{Potts model and spin clusters}
In the following we study the ferromagnetic Potts model whose Hamiltonian is written as:
\begin{equation}
{\cal H} = -J \sum_{\langle ij\rangle} \delta_{S_i S_j}   \; ,
\end{equation}
where the sum is over all nearest neighbors pairs of spins of a $d$-dimensional system, $J>0$ and $\delta$ the Kronecker
delta and $S_i$ the spin at site $i$ taking integer values in $[1,Q]$. The $Q\to 1$ limit of this model is known to correspond to percolation.
This model undergoes a magnetic phase transition at the inverse temperature $\beta_c=1/T_c$ and this transition is
continuous for $Q\le Q_c(d)$. For $d=2$, $Q_c(2)=4$ and $\beta_c=\ln(1+\sqrt{Q})$ for the square lattice~\cite{wu_potts_1982}.

In the study of the Potts model two types of clusters emerge: the FK clusters
and the same spin clusters. The FK clusters appear naturally when the Potts
partition function is rewritten in the form of the random cluster model~\cite{fortuin_random-cluster_1972}. The
construction of FK clusters is simple, edges between spins with the same value are open with
probability $p=1-e^{-\beta J}$ and edges between spins with different values are
always closed. A FK cluster is a connected component of open edges. Defined
this way the Potts model is really a correlated site-bond percolation model. The FK clusters undergo a continuous
percolation transition at $T_c$ in $d = 2$ for $Q\le 4$ which coincides with the magnetic transition and their exponents
are the usual Potts model ones.

We can also define
another type of clusters by the same construction as before with $p=1$,
\textit{i. e.} all bonds connecting spins with equal value are open. We will call the connected
components of such a construction same-spin clusters or simply spin clusters. Those
clusters are the one occurring naturally in experimental realisations of the
Potts model and are most frequently studied in coarsening and more
generally out of equilibrium situations~\cite{blanchard_frozen_2013,blanchard_morphological_2012,blanchard_how_2013}.
The spin clusters also undergo a continuous percolation transition but at a temperature $T_s$ in general different from
$T_c$. It is only in $d=2$ that $T_s=T_c$, hence both types of clusters are self-invariant at the critical temperature. 
Moreover, in this case, they are both described by a conformal field theory (CFT) with a single central charge $c$ . However they correspond to
different realisations of the CFT since their critical exponents differ, hence the conformal dimensions of the operators of the
two models are different as well.
As mentioned in the introduction, it has been shown that the Ising spin clusters arise naturally if one considers the $Q=1$ Potts model with
dilution at its tricritical point~\cite{coniglio_clusters_1980}. This correspondence has been used extensively, for example to derive the fractal dimension of
the Ising spin clusters~\cite{stella_scaling_1989}. 
 



Two dimensional critical systems can be conveniently described using the Coulomb gas
formalism~\cite{francesco_relations_1987,nienhuis_coulomb_1987}. In the case of the Potts model, the Coulomb gas parameter $g$ can be parametrized in the following way:
\begin{eqnarray}
Q&=&2+2\cos\left(\frac{\pi g}{2}\right).
\label{g}
\end{eqnarray}
For the critical branch we have $g\in[2,4]$ and $g\in[4,8]$ for the tricritical branch following the conjectured duality.
In particular we have $g=8/3$ for percolation, $g=3$ for Ising FK clusters and $g=16/3$ for Ising spin clusters. 

\section{Wrapping probabilities on a torus\label{sec_wrapp}}
In this section we recall shortly the derivation of FK clusters wrapping probabilities on a torus. For more details, we
refer the reader to the works~\cite{francesco_relations_1987,pinson_critical_1994,arguin_homology_2002}.  We will
consider a torus $\mathbb{T}=\mathbb{C}/(\mathbb{Z}+\tau\mathbb{Z})$ with parameter $\tau\in \mathbb{C}$ such that
$\tau=\tau_{\scriptscriptstyle R}+i\tau_{\scriptscriptstyle I}$. In other words, we build a torus in the complex plane
by identifying the opposite sides of the parallelogram with vertices $0$, $1$, $1+\tau$, $\tau$. The first homology
group of a torus is $H_1(\mathbb{T})=\mathbb{Z}\times\mathbb{Z}$. We will be interested in the probabilities $\pi^{a,b}(\tau)$ of
subgroups $\{a,b\}$ of $\mathbb{Z}\times\mathbb{Z}$ with $(a,b)\in \mathbb{Z}^2$, $\pi^0(\tau)$ for the trivial subgroup
$\{0\}$ and $\pi^{+}(\tau)$ for $\mathbb{Z}\times\mathbb{Z}$. The configurations corresponding to $\{a,b\}$ contain at
least one non-trivial cluster winding $a$ times horizontally from the left to the right and $b$ times vertically from
top to bottom. It can be easily shown that if a configuration contains a non trivial $\{a,b\}$ cluster then all other
non trivial clusters that might be present are of the same topology~\cite{francesco_relations_1987}. Since the cluster
boundaries cannot intersect themselves $a$ and $b$ must be coprime~\cite{francesco_relations_1987}. The clusters
$\{-a,-b\}$ are identified with $\{a,b\}$ so we assume $a>0$. The subgroup $\mathbb{Z}\times\mathbb{Z}$ is generated by
cross-shaped clusters winding in both direction around the torus. It is easy to check that a given configuration can
only contain one such cluster.

From Eq.~(\ref{prob}) we see that the derivation reduces to the computation of restricted partition functions. In
particular, the partition function $Z_Q$ of the $Q$ states Potts model can be decomposed on the different cluster
homology classes such that  $Z_Q=Z^0_Q+Z^{+}_Q+\sum_{a\wedge b=1}Z^{a,b}_Q$ where $a\wedge b$ stands for the greatest
common divisor of $a$ and $b$. The sum is taken with $a$ and $b$ coprime for the reason mentioned previously.  $Z^0_Q$,
$Z^{+}_Q$ and $Z^{a,b}_Q$ are the partition functions restricted to the configurations containing respectively only
trivial clusters, one cross cluster and a least one non trivial cluster of type $\{a,b\}$. 

The derivation of the crossing probabilities uses extensively a chain of equivalences between the $Q$ states Potts
model, the six vertex model and solid on solid (SOS) models which are known to renormalise into a bosonic free field with parameter $g$.
We will not describe those equivalences which have been extensively discussed in the literature, \textit{e.g.}
see~\cite{baxter_exactly_2008}. We just need to know that clusters are represented by oriented loops formed by
their interfaces and that height variables are introduced such that the height varies of $\pm \pi/2$ when crossing a
line depending on its orientation. It is this height which becomes the bosonic free field in the continuum limit. One
can easily see that a configuration containing only trivial clusters yields a periodic field in both periods of the
torus. On the contrary, a non trivial cluster induces discontinuities in the field, for example a single $\{a,b\}$ cluster induces a 
discontinuity of $b\pi$ horizontally and $a\pi$ vertically because for a single non trivial cluster there is a pair of interfaces.

Di~Francesco~\textit{et al.}~\cite{francesco_relations_1987} showed that the partition function $Z_Q$ can be expressed
with bosonic partition functions $Z_{m,m'}(g)$ corresponding to a bosonic free field with discontinuities $m\pi$ horizontally and
$m'\pi$ vertically. $Z_{m,m'}(g)$ has the following expression:
\begin{equation}
Z_{m,m'}(g)=\sqrt{\frac{g}{\tau_I}}\frac{1}{|\eta(q)|^2}\exp\left[-\pi g\frac{m^2\tau_I^2 + (m'-m\tau_R)^2}{\tau_I}\right]
\label{zboson}
\end{equation}
with $q=e^{2i\pi\tau}$ and $\eta(q)=q^{1/24}\prod_{n=1}^{\infty}(1-q^n)$ is the Dedekind $\eta$ function.
One has to be careful so as to give the correct weight $\sqrt{Q}$ to non trivial interfaces. To achieve this we have to add an
electric charge $e_0$ such that:
\begin{equation}
\sqrt{Q}=2\cos\left(\frac{\pi e_0}{2}\right).
\label{charge}
\end{equation}
With this we can give the proper weight to all interfaces in the partition function by multiplying $Z_{m,m'}(g)$ to a
cosine factor $\cos\left[\pi e_0 (m\wedge m')\right]$.
One can see that choosing $e_0=1$ kills the contribution of non trivial loops. This allows the determination of
$Z^0_Q$:
\begin{equation}
Z^0_Q=\frac{1}{2}\sum_{m,m'\in \mathbb{Z}}Z_{m,m'}(g/4)\cos\left[\pi (m\wedge m')\right]=\frac{1}{2}Z_{c}(g,1)
\end{equation}
where $Z_c(g,e_0)$ is the coulombic partition function defined by Di~Francesco~\textit{et al.} whose expression is:
\begin{equation}
Z_c(g,e_0)=\sum_{m,m'\in \mathbb{Z}}Z_{m,m'}(g/4)\cos\left[\pi e_0 (m\wedge m')\right].
\end{equation}
Then using the duality relation $Z^+_Q=QZ^0_Q$ coming from Euler's relation, one obtains the partition function restricted to cross-shaped clusters.
Now to obtain $Z^{a,b}_Q$ one has to consider the sum over all configurations with the desired discontinuities while
removing the undesired contribution of cross-shaped clusters giving the same discontinuities:
\begin{equation}
Z^{a,b}_Q=\sum_{\substack{m=bk\\m'=ak\\k\in \mathbb{Z}\backslash\{0\}}}\left(Z_{m,m'}(g/4)\cos\left[e_0\pi
(m\wedge m')\right]-Z_{m,m'}(g/4)\cos\left[\pi (m\wedge m')\right]\right).
\end{equation}

The complete partition function $Z_Q$ can be obtained by summing over all topology classes:
\begin{equation}
 Z_Q  =\frac{(Q+1)}{2}Z_c(g,1)+\sum_{a\wedge b=1}Z^{a,b}_Q=\frac{(Q-1)}{2}Z_c(g,1)+Z_c(g,e_0).
\end{equation}

\section{Closed forms for Ising spin clusters probabilities}
We now turn to the determination of the closed form for the probability $\pi_{2,\mathrm{spin}}^{a,b}(\tau)$ for Ising
spin clusters to wind around the torus. To compute this probability, we use the equivalence to the $Q'=1$ Potts model
with dilution at its tricritical point, so that $\pi^{\mathrm{spin}}_2(\{a,b\})=\pi^{\mathrm{tri}}_1(\{a,b\})$.  We know
from Eq.~(\ref{g}) that the tricritical Potts model with $Q'=1$ renormalises to a Coulomb gas with parameter $g=16/3$. The value $e_0=2/3$ comes from
Eq.~(\ref{charge}) taking $Q'=1$. We can now compute $Z^{a,b}_{2,\mathrm{spin}}$:
\begin{eqnarray}
Z^{a,b}_{2,\mathrm{spin}}&=&\sum_{\substack{m=bk\\m'=ak\\k\in \mathbb{Z}\backslash\{0\}}}\left[Z_{m,m'}(4/3)\cos\left[2\pi/3  (m\wedge m')\right]-Z_{m,m'}(4/3)\cos\left[\pi (m\wedge m')\right]\right]\nonumber\\
		      &=&\sum_{k\in \mathbb{Z}\backslash\{0\}}\left[\cos\left(\frac{2\pi}{3}k\right)-(-1)^k\right]Z_{bk,ak}(4/3)\nonumber\\
		      &=&\frac{2/\sqrt{3}}{{\tau_I}^{1/2}|\eta(q)|^2}\left[\sum_{k\in \mathbb{Z}\backslash\{0\}}\cos\left(\frac{2\pi}{3}k\right)e^{-\pi\frac{4}{3}\frac{|\tau_{a,b}|^2}{\tau_I}k^2}-\sum_{k\in \mathbb{Z}\backslash\{0\}}(-1)^ke^{-\pi\frac{4}{3}\frac{|\tau_{a,b}|^2}{\tau_I}k^2}\right]\nonumber\\
\end{eqnarray}
where $\tau_{a,b}=a-b\tau$.  The second sum can be expressed with a Jacobi $\theta$ function as
$\theta_4\left(i\frac{4}{3}\frac{|\tau_{a,b}|^2}{\tau_I}\right)$ with $\theta_4(\tau)=\sum_{n\in\mathbb{Z}}(-1)^n
q^{n^2/2}$ with $q=e^{2i\pi\tau}$. The first sum has to be decomposed so that the cosine has a constant value,
\textit{i. e.} $k\equiv 0,1,2\bmod{3}$. Then the sums obtained are all expressed with Jacobi $\theta$ functions:
\begin{eqnarray}
Z^{a,b}_{2,\mathrm{spin}}(\tau)  &=&\frac{2/\sqrt{3}}{{\tau_I}^{1/2}|\eta(q)|^2}\left[\left(\sum_{k\equiv 0 \bmod{3}}-\frac{1}{2}\sum_{k\equiv 1,2\bmod{3}}\right)e^{-\pi\frac{4}{3}\frac{|\tau_{a,b}|^2}{\tau_I}k^2}-\theta_4\left(i \frac{4}{3}\frac{|\tau_{a,b}|^2}{\tau_I}\right)\right]\nonumber\\
		   &=&\frac{2/\sqrt{3}}{{\tau_I}^{1/2}|\eta(q)|^2}\left[\frac{3}{2}\theta_3\left(12i\frac{|\tau_{a,b}|^2}{\tau_I}\right)
								       -\frac{1}{2}\theta_3\left(i \frac{4}{3}\frac{|\tau_{a,b}|^2}{\tau_I}\right)
								       -\theta_4\left(i \frac{4}{3}\frac{|\tau_{a,b}|^2}{\tau_I}\right)\right]\nonumber\\
\end{eqnarray}
where $\theta_2(\tau)=\sum_{n\in\mathbb{Z}}q^{(n+1/2)^2/2}$ and $\theta_3(\tau)=\sum_{n\in\mathbb{Z}}q^{n^2/2}$.  We now
use identities on the $\theta$ functions to inverse their arguments to obtain this compact expression:
\begin{eqnarray}
Z^{a,b}_{2,\mathrm{spin}}(\tau)&=&\frac{1}{2|\tau_{a,b}||\eta(q)|^2}\left[\theta_3\left(i\frac{\tau_I}{12|\tau_{a,b}|^2}\right)-\theta_3\left(i\frac{3}{4}\frac{\tau_I}{|\tau_{a,b}|^2}\right)-2\theta_2\left(i\frac{3}{4}\frac{\tau_I}{|\tau_{a,b}|^2}\right)\right].\nonumber\\
\end{eqnarray}
With the expression of the Ising model partition function on a torus~\cite{francesco_conformal_1997} one obtains the probability
$\pi_{2,\mathrm{spin}}^{a,b}(\tau)$:
\begin{equation}
	\pi_{2,\mathrm{spin}}^{a,b}(\tau)=\frac{1}{|\tau_{a,b}||\eta(q)|}\frac{\theta_3\left(i\frac{\tau_I}{12|\tau_{a,b}|^2}\right)-\theta_3\left(i\frac{3}{4}\frac{\tau_I}{|\tau_{a,b}|^2}\right)-2\theta_2\left(i\frac{3}{4}\frac{\tau_I}{|\tau_{a,b}|^2}\right)}{|\theta_2(\tau)|+|\theta_3(\tau)|+|\theta_4(\tau)|}
\label{prob_ising}
\end{equation}



\section{Numerical check}
We have checked numerically by means of Monte Carlo simulations the formula~(\ref{prob_ising}) obtained for spin clusters.
The systems were equilibrated at $T_c$ using a Wolff cluster algorithm~\cite{wolff_collective_1989}. $10^6$ independent
configurations were simulated for each value of the torus parameter $\tau$ considered. 
In all the simulations we used the triangular lattice where the interfaces are defined with no ambiguity as opposed to
the square lattice where there exists intersecting interfaces. 
We have simulated the common case of untwisted tori such that $\tau=i r$ for different aspect
ratio $r$. To implement those tori one has to, starting from a square lattice, add bonds on alternate diagonals on even
or odd row. Adding periodic boundary conditions one obtains the desired tori with $\tau=i r$.

\begin{figure}[h!]
	 \begin{center}
\includegraphics[scale=1.45]{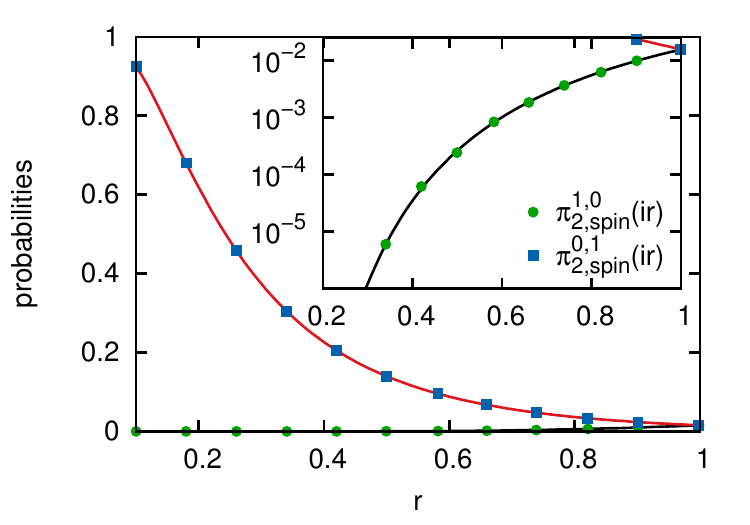}
	 \end{center}
 \caption{(color online) Ising spin clusters wrapping probability $\pi_{2,\mathrm{spin}}^{0,1}(ir)$ (blue
squares and red line) and $\pi_{2,\mathrm{spin}}^{1,0}(ir)$ (green circles and black line) at $T_c$ vs. the aspect ratio
$r$. The lines correspond to
the expression of Eq.~(\ref{prob_ising}) and the symbols correspond to the results of simulations. The error bars are
smaller than the symbols. In the inset the same probabilities are represented with the ordinate in logarithmic scale.}
 \label{pbc_1}
 \end{figure}
\begin{figure}[h!]
	 \begin{center}
\includegraphics[scale=1.45]{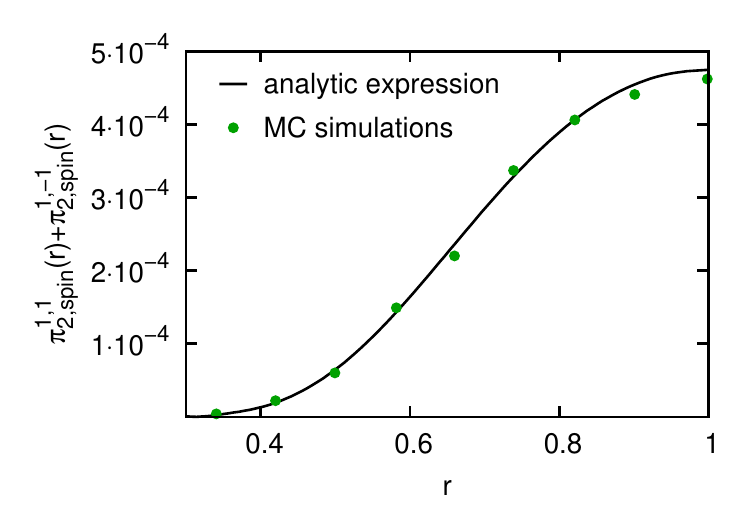}
	 \end{center}
 \caption{(color online) Ising spin clusters wrapping probability
$\pi_{2,\mathrm{spin}}^{1,1}(ir)+\pi_{2,\mathrm{spin}}^{1,-1}(ir)$ (green circles and black line) at $T_c$ vs. the aspect ratio
$r$. The line corresponds to the expression of Eq.~(\ref{prob_ising}) and the symbols correspond to the results of simulations.}
 \label{pbc_2}
 \end{figure}

We present the results for this case ($\tau=ir$) with $r\in[0.1,1]$. For a system composed of $L_x$ sites
horizontally and $L_y$ sites vertically $r=(\sqrt3L_y/2)/L_x$. The factor $\sqrt3/2$ accounts for the fact that the
system is composed of hexagons instead of squares. In the simulations we took $L_x=128$ and
$L_y\in\{148,164,180,200,224,254,296,352,434,568,822,1478\}$. This gives aspect ratios $r\in[1,10]$. We then use the
trivial symmetry $\pi_{2,\mathrm{spin}}^{0,1}(i/r)=\pi_{2,\mathrm{spin}}^{1,0}(ir)$ 
to be able to plot graphs with $r\in[0,1]$.
In Fig.~\ref{pbc_1}, $\pi_{2,\mathrm{spin}}^{0,1}(ir)$ and $\pi_{2,\mathrm{spin}}^{1,0}(ir)$ obtained 
from Eq.~(\ref{prob_ising}) are represented as function of $r$ (lines) and compared with the probabilities measured from
Monte Carlo simulations (symbols). 
 The agreement is excellent. Since in this range of aspect ratios $\pi_{2,\mathrm{spin}}^{1,0}(ir)$ is at most of order
 $10^{-2}$, hence barely visible, we add an inset with the ordinate in logarithmic scale.  
In Fig.~\ref{pbc_2}, the probabilities for clusters winding once vertically and horizontally are presented.
The sum $\pi_{2,\mathrm{spin}}^{1,1}(ir)+\pi_{2,\mathrm{spin}}^{1,-1}(ir)$ is shown instead of each one individually
because of the symmetry relation $\pi_{2,\mathrm{spin}}^{1,1}(ir)=\pi_{2,\mathrm{spin}}^{1,-1}(ir)$. Even at its maximum
at $r=1$ the sum is of the order of $5.10^{-5}$ which makes its numerical estimation less accurate than before but the
agreement with the formula is still good. The probabilities for other homotopy groups are too small to allow a numerical
estimation. 

\section{Wrapping probabilities for cross clusters on untwisted tori}
In this section we are interested in clusters contributing to the subgroup $\mathbb{Z}\times\mathbb{Z}$, \textit{i.e.}
cross-shaped clusters.  In~\cite{ziff_shape-dependent_1999} Ziff~\textit{et. al.} obtained a closed form for cross
clusters on untwisted tori ($\tau=i r$).  They give a closed form only for percolation but it is possible to do the same
for the FK and spin clusters as well. As seen in section~\ref{sec_wrapp}, the partition function of interest $Z^+_Q$ is:
\begin{eqnarray}
Z^+_Q&=&\frac{Q}{2}Z_c(g,1)=\frac{Q}{2}\sum_{m,m'\in \mathbb{Z}}\left[Z_{m,m'}(g)-Z_{m,m'}(g/4)\right]\nonumber \\
&=&\frac{Q}{2}\left[Z_c(4g,0)-Z_c(g,0)\right].
\end{eqnarray}
For untwisted tori ($\tau=i r$) $Z_c(g,0)$ becomes:
\begin{eqnarray}
	Z_c(g,0)&=&\sqrt{\frac{g}{r}}\frac{1}{|\eta(q)|^2}\sum_{m,m'\in \mathbb{Z}}\exp\left[-\pi g\left(m^2r +\frac{m'^2}{r}\right)\right]\nonumber \\
	      &=&\sqrt{\frac{g}{r}}\frac{\theta_3\left(igr\right)\theta_3\left(\frac{ig}{r}\right)}{|\eta(q)|^2}\nonumber \\
\end{eqnarray}
because in this case the sum over two integers reduces to a product of $\theta$ functions.
Finally:
\begin{eqnarray}
Z^+_Q &=&\frac{Q}{4}\sqrt{\frac{g}{r}}\frac{2\theta_3\left(igr\right)\theta_3\left(\frac{ig}{r}\right)-\theta_3\left(\frac{igr}{4}\right)\theta_3\left(\frac{ig}{4r}\right)}{|\eta(q)|^2}\nonumber \\
\end{eqnarray}
This formula is valid for all $g$ but the probability is in general more complicated because of the total partition
function. For the Ising model the probability is simply:
\begin{equation}
\pi^+_Q(r)=\frac{Q}{2|\eta(q)|}\sqrt{\frac{g}{r}}\frac{2\theta_3\left(igr\right)\theta_3\left(\frac{ig}{r}\right)-\theta_3\left(\frac{igr}{4}\right)\theta_3\left(\frac{ig}{4r}\right)}{|\theta_2\left(ir\right)|+|\theta_3\left(ir\right)|+|\theta_4\left(ir\right)|}.
\label{cross_ising}
\end{equation}
Taking $g=3$ and $Q=2$ yields the result for the Ising FK clusters in agreement with the values obtained by Arguin. For
spin clusters the probability is expected to be given by taking $g=16/3$ and $Q=1$. The situation is however slightly
more subtle as a direct measurement from MC simulations shows that $\pi_{2,\mathrm{spin}}^0=0$ and that
$\pi_{2,\mathrm{spin}}^+$ is twice the quantity given by Eq.~(\ref{cross_ising}). This can be explained by the fact that
when spin clusters of a given color are all trivial then there exists a cluster of the other color wrapping in both
direction and thus contributing to the subgroup $\mathbb{Z}\times\mathbb{Z}$. This means that no configuration can
contribute to $\{0\}$ while considering simultaneously both colors. Furthermore one has to remember that we are
extrapolating those results from the critical line using the mapping to a tricritical model with only one state. By doing
so, one of the two spin values of the Ising spin is absorbed to generate the vacancies of the tricritical model. Hence
considering one color at a time is natural when considering the properties of spin clusters. It makes no difference to the
discussion for the configurations contributing to the subgroups $\{a,b\}$ since in this case at least one cluster of each
color is present. However, for the cross-shaped clusters, it is necessary to consider only one color as those clusters
come alone. This enables us to recover the relation $\pi_{2,\mathrm{spin}}^0=\pi_{2,\mathrm{spin}}^+$ expected with
$Q'=1$. Indeed, because of the symmetry between the two colors, considering only one color reduces the value $\pi_{2,\mathrm{spin}}^+$ measured in
MC simulation by a factor two which is in agreement with
Eq.~(\ref{cross_ising}) with $g=16/3$ and $Q=1$ .

\section*{Conclusion}  
In this work, we checked that the natural analytic continuation from the critical branch yields the correct results for
the wrapping probabilities of Ising spin clusters. We also obtained an elegant expression for those wrapping
probabilities. As a generalisation the study of the 3-Potts spin clusters is definitely interesting but might be more
subtle as the mapping corresponds to a tricritical Potts model with a non integer number of states. This case is also
interesting since the conjectured continuation to the tricritical branch is known to fail in some cases as already
mentioned. This will be treated in another work.  

\section*{Acknowledgments}
I thank M.~Picco and R.~Santachiara for useful discussions on the content of this article and M.~Picco for his careful
reading of the manuscript. \\

\bibliography{spin_torus.bib}

\end{document}